# Identification of Two Distinct Antiferromagnetic Phases in the Au-Al-Gd Quasicrystal Approximant

Farid Labib[1], Takanori Sugimoto[2,3,4], Hiroyuki Takakura[5], Asuka Ishikawa[1] and Ryuji Tamura[6]

[1] *Research Institute of Science and Technology, Tokyo University of Science, Tokyo 125-8585, Japan*
[2] *Center for Quantum Information and Quantum Biology, Osaka University, Toyonaka, Osaka 560-8531, Japan*
[3] *Advanced Science Research Center, Japan Atomic Energy Agency, Tokai, Ibaraki 319-1195, Japan*
[4] *Computational Materials Science Research Team, Riken Center for Computational Science (R-CCS), Kobe, Hyogo 650-0047, Japan*
[5] *Division of Applied Physics, Faculty of Engineering, Hokkaido University, Sapporo 060-8628, Japan*
[6] *Department of Materials Science and Technology, Tokyo University of Science, Tokyo 125-8585, Japan*

The structural and physical properties of two single-grain Au-Al-Gd 1/1 approximant crystals (ACs) with analyzed compositions $Au_{73.67}Al_{12.24}Gd_{14.09}$ and $Au_{74.67}Al_{11.35}Gd_{13.98}$ were thoroughly investigated. The two variants are iso-structural (cubic, space group: $Im\bar{3}$), undergoing sharp antiferromagnetic (AFM) transitions at $T_N$ = 9.6 K and $T_N$ = 8.3 K, respectively. Specific heat measurements in both samples evidenced sharp jump at $T_N$, associated with the AFM transition, accompanied by broad Schottky-type anomalies at lower temperatures. Strikingly, the two 1/1 ACs exhibited markedly different responses to applied magnetic fields: one displayed a pronounced metamagnetic anomaly while the other did not. In this regard, variational calculations with adiabatic transformation of the magnetic field in realistic structural models identified the existence of two possible AFM phases: the cuboc phase with a metamagnetic anomaly and the cluster Néel phase without it, thereby clearly revealing the possible magnetic structures of the present AFM variants. These findings have led to the establishment of a comprehensive magnetic phase diagram for Heisenberg-type 1/1 ACs. This work not only advances our understanding of magnetic phase transitions in these complex systems but also suggests the existence of a broader spectrum of unexplored magnetic states in Tsai-type materials.

## I. INTRODUCTION

Tsai-type icosahedral quasicrystals (iQCs) and their cubic approximant crystals (ACs) represent an emerging class of strongly correlated electron systems. Their atomic structure commonly comprises a common rhombic triacontahedron (RTH) unit arranged aperiodically (periodically in the case of ACs) in a physical space [1]. Figure 1(a) shows the arrangement of RTH units within a unit cell of 1/1 AC whose lattice parameter (*a*) is typically within 13 – 15 Å. The RTH unit itself, as schematically illustrated in Fig. 1(b), accommodates multiple concentric inner shells being (from the outermost one): an icosidodecahedron, an icosahedron and a dodecahedron, all surrounding a central tetrahedron unit which is often orientationally disordered [2] but occasionally becomes ordered [3] depending on the rare-earth (RE) type.

In 1/1 AC, each unit cell contains approximately 162 atoms, of which 24 are RE elements positioned at the vertices of an icosahedron [see Fig. 1(b)]. The rest of the atomic sites are typically occupied by combinations of d- and p-block elements from the periodic table. Magnetic properties of these compounds are primarily governed by indirect exchange interactions between the localized RE ions, mediated by conduction electrons through the Ruderman-Kittel-Kasuya-Yosida (RKKY) mechanism, often resulting in spin-glass (SG)-like frustrated behavior.

Whether aperiodic structures of iQCs could host long-range magnetic orders has been a debating issue for decades until recently that ferromagnetic (FM) [4,5] and antiferromagnetic (AFM) [6] orders were confirmed in real iQCs. Additionally, a

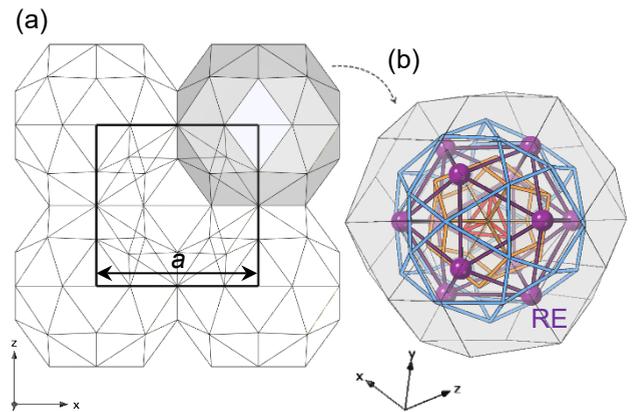

**Fig. 1.** (a) Arrangement of RTH clusters within a unit cell of Tsai-type 1/1 AC. (b) The RTH unit has several inner shells; from the outermost one: an icosidodecahedron (blue cage), an icosahedron (pink cage) encompassing rare-earth (RE) elements, a dodecahedron (orange cage), and a central tetrahedron (unit in red). The vertices of the icosahedron shell are exclusively occupied by RE elements.

…

universal rule has been revealed that reducing the electron-per-atom ($e/a$) ratio in Tsai-type ACs through compositional tuning enables the stabilization of long-range FM [4,7–13] and AFM [14–20] orders. Here, the $e/a$ refers to the ratio of the total number of valence electrons of the constituent elements divided by the total number of atoms present in the compound.

Moreover, neutron diffraction experiments on several Tb, Dy, and Ho-containing AFM 1/1 ACs have identified a unique noncoplanar magnetic structure [17,21,22] characterized by a whirling spin configuration around the crystallographic [111] axis. For Heisenberg Tsai-type 1/1 ACs containing Gd, however, no experimental report about their magnetic structure exists to date. This is primarily due to difficulties associated with collecting neutron diffraction owing to high neutron absorption cross-sections of Gd.

This study reports the synthesis, structural analysis, and magnetic properties of two distinct AFM variants of Au-Al-Gd 1/1 ACs in a single crystal form in average diameter of ~ 5 mm. Despite their structural similarity, the two variants showcase different magnetic behaviors under external magnetic fields, with only one displaying a field-induced metamagnetic anomaly. Additionally, variational calculations performed on real structural models identified two possible AFM phases: the *cuboc* phase with a metamagnetic anomaly in field dependence magnetization curve and the *cluster Néel* phase without it. These results being in qualitative agreement with the experimental results provide insight into the possible magnetic structures of the present AFM 1/1 ACs. Based on the results, a comprehensive magnetic phase diagram for Heisenberg 1/1 AC systems is developed.

## II. EXPERIMENT

Single crystals of Au-Al-Gd 1/1 ACs are prepared by encapsulating high purity Au (99.99%), Al (99.999%) and Gd (99.9%) elements inside alumina crucibles (10 mm in diameter and 5 mm in length). For that purpose, steel grids are affixed over the raw elements inside the crucibles, followed by wrapping Ta foil at the top of crucibles and sealing them inside quartz tubes under a depressurized Ar gas to prevent elemental oxidation. For the synthesis, a solution growth technique is applied, based on which the prepared capsules were heated up to 1273 K and halted at that temperature for 10 hours for solutionizing. The temperature then cooled down to 1033 K at a rate of 1K/hour, at which temperature the Au-rich solution was decanted by implementing a centrifuge apparatus resulting in single crystals in average diameter of ~ 5 mm.

The grown single crystals were examined by single crystal x-ray diffraction (SCXRD) at ambient temperature using an XtaLAB Synergy-R single-crystal diffractometer equipped with Hybrid Pixel Array Detector (HyPix6000, Rigaku) with Mo K$\alpha$ radiation ($\lambda$ = 0.71073 Å). For microstructure observation, a scanning electron microscopy (SEM); JEOL JSM-IT100 equipped with energy dispersive x-ray (EDX) spectrometer was utilized. For bulk magnetization measurement, superconducting quantum interference device (SQUID) magnetometer (Quantum Design, MPMS3) was utilized under zero-field-cooled (ZFC) and field-cooled (FC) modes within a temperature range of 1.8 K to 300 K and under external dc fields up to $7 \times 10^4$ Oe. Additionally, ac magnetic susceptibility measurements were carried out at frequencies ranging from 0.1 Hz to 100 Hz within a temperature range of 2 – 35 K and ac magnetic field amplitude ($H_{ac}$) of 1 Oe. To prevent signal saturation during magnetization measurements and to examine the crystallographic orientation dependence, the single crystals were carefully polished along the [100] or [110] facets into equidimensional 1×1×1 mm cubes, with each face corresponding to [100] or [110], depending on the sample. A representative optical microscopy image of the polished cube studied in this work is provided in Figure S1.

Specific heat measurements were conducted in a temperature range of 2 – 40 K by a thermal relaxation method using a Quantum Design Physical Property Measurement System (PPMS). Furthermore, based on an effective model of classical spins, a variational method starting with zero magnetic-field ground state was performed to understand the experimental results leading to magnetization curves at zero temperature.

## III. RESULTS

Figure 2(a) and (d) display optical microscopy images of two synthesized single crystals with slightly different Au concentrations. The starting compositions for the solution growth were $Au_{76.0}Al_{16.0}Gd_{8.0}$ for (a) and $Au_{77.4}Al_{14.6}Gd_{8.0}$ for (d). The average compositions obtained from EDX elemental analysis from four distinct spots on the polished surfaces are $Au_{73.67}Al_{12.24}Gd_{14.09}$ and $Au_{74.67}Al_{11.35}Gd_{13.98}$, respectively. Corresponding SEM backscattering electron images of the polished surfaces and detailed EDX elemental analysis results are provided in Fig. S2 and Table S1 of the Supplemental Material. For simplicity, the single crystals with lower and higher Au concentrations are denoted as AAG-I and AAG-II, respectively, throughout the manuscript. As shown in Fig. 2(a) and (d), the single crystals exhibit natural facets corresponding to the [110] and [100] planes, confirmed by their respective Laue x-ray backscattering patterns in Fig. 2(b) and (e). For comparison, calculated Laue x-ray patterns along the [110] and

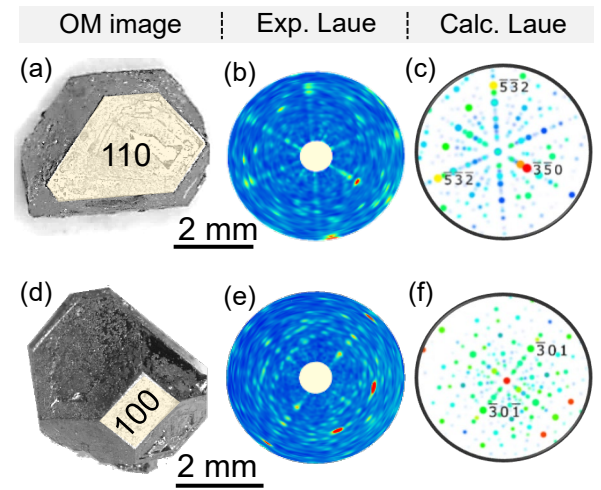

**Fig. 2.** The optical microscopy images from single crystals with average diameters of ~ 5 mm synthesized via solution growth technique from starting compositions of (a) $Au_{76.0}Al_{16.0}Gd_{8.0}$ and (d) $Au_{77.4}Al_{14.6}Gd_{8.0}$. (b,e) Experimental Laue x-ray diffraction patterns from [110] and [100] facets, highlighted in (a) and (d). (c,f) calculated Laue x-ray backscattering patterns along [110] and [100] directions using atomic structure parameters refined in this work.

…

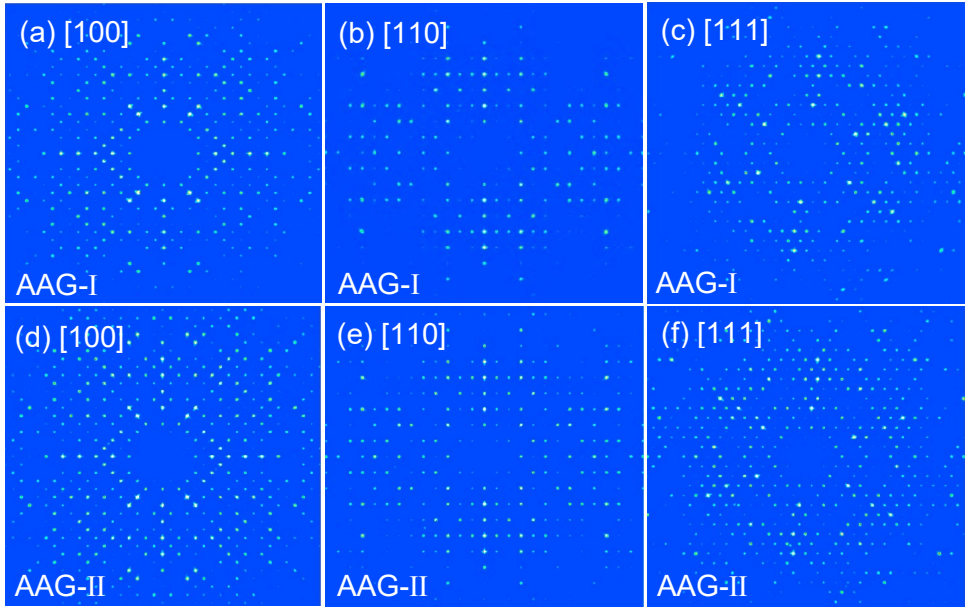

**Fig. 3.** Reconstructed reciprocal space sections of the SCXRD datasets obtained from single crystals of (a-c) AAG-I and (d-f) AAG-II.

[100] directions using the refined structural models of AAG-I and AAG-II, are provided in Fig. 2(c) and (f), respectively.

For structural characterization, the single crystals are examined by SCXRD experiment. Reciprocal space sections perpendicular to the primary zone axes of [100], [110], and [111] were generated from the diffraction data, as illustrated in Fig. 3(a-c) for AAG-I and Fig. 3(d-f) for AAG-II. The CrysAlisPRO software suite (Rigaku Oxford Diffraction, 2018) was employed for indexing, integration of peak intensities, and absorption correction. The diffraction patterns in Fig. 3 confirm that both samples are 1/1 ACs of the Tsai-type category, exhibiting a cubic lattice with the space group $Im\bar{3}$ with no sign of systematic extinction rule violation. No difference is noticed in the symmetry of the two 1/1 AC variants.

To refine atomic structures, initial models were obtained using SHELXT [23] followed by subsequent structure refinements using SHELXL [24]. The experimental conditions, refinement parameters and details of the final model including atomic coordinates, Wyckoff positions, site occupations, and equivalent isotropic displacement parameters ($U_{eq.}$) are provided in Tables S(2–5) of the Supplemental Material. Figure 4 presents refined structural model of the AAG-II, viewed along the [100] axis. The structural model corresponding to the AAG-I is provided separately in Fig. S3 of the Supplemental Material. In the atomic structure representation, Au, Al, Gd atoms and unoccupied sites are represented by dark yellow, light blue, violet and white spheres, respectively. In both structures, the main building unit is an RTH cluster. The lattice parameter of the AAG-II is 14.8187(1) Å being slightly larger than 14.8085(1) Å in the AAG-I, reflecting lattice expansion due to slightly higher Au content in the structure. Notice that the centers of the cube-shaped interstices connecting adjacent dodecahedra along 3-fold axes [see Fig. 4(b)] are fully occupied by Al. The next preferred site for Al is the vertices of the RTH shell [see Fig. 4(e)]. As a characteristic feature in QCs and ACs, the ratio of the second nearest, $d$ in Fig. 4(c), to nearest neighbor, $a$, $b$, and $c$, amounts to ~ 1.6196 – 1.6360, close to the golden mean defined by $\tau = (1+\sqrt{5})/2 = 1.6180$. Therefore, except minor dissimilarities, the two 1/1 ACs could be considered as isostructural.

Next, physical properties of the two variants are evaluated beginning with their temperature dependence of the dc magnetic susceptibility ($M/H$) under FC and ZFC conditions within a temperature range of 1.8 – 30 K, as shown in the main panel of Figs. 5(a) and (b). The corresponding high-temperature inverse magnetic susceptibility ($H/M$) of the samples within a temperature range of 1.8 – 300 K is presented in Fig. S4 of the Supplemental Material. Note that the magnetic susceptibility of the samples was measured separately with the applied magnetic field ($H$) aligned along the principal crystallographic axes [100], [110], and [111]. However, no magnetic anisotropy was observed across these orientations. The data shown in Fig. 5 and Fig. S4 correspond to measurements taken with the magnetic field applied parallel to the [100] axis. The $H/M$ curves in both ACs exhibit linear behavior fitting well to the Curie-Weiss law: $\chi(T) = N_A \mu_{\text{eff}}^2 \mu_B^2 / 3k_B(T - \theta_w) + \chi_0$, where $N_A$, $\mu_{\text{eff}}$, $\mu_B$, $k_B$, $\theta_w$, and $\chi_0$ denote the Avogadro number, effective magnetic moment, Bohr magneton, Boltzmann constant, Curie-Weiss temperature, and the temperature-independent magnetic susceptibility, respectively. By extrapolating linear least-square fittings within a temperature range of 100 K < $T$ < 300 K, $\theta_w$ values of +11.1(2) K and +9.2(2) K are derived for the AAG-I and AAG-II 1/1 ACs, respectively, suggesting a dominant FM component (though slightly weaker in the latter) in both samples. The $\mu_{\text{eff}}$ values of both ACs are within a range of 7.85 – 8.12 $\mu_B$, close to 7.94 $\mu_B$, i.e., the calculated value for free $Gd^{3+}$ ions defined as $g_J(J(J+1))^{0.5}$ $\mu_B$ [25]. This indicates the localization of the magnetic moments on $Gd^{3+}$ sites.

At low temperatures, as shown in Fig. 5(a) and (b), both AAG-I and AAG-II exhibit sharp cusps in the FC and ZFC susceptibilities below $T_N$ = 9.6 K and 8.3 K, respectively, corresponding to the onset of AFM order. The slightly lower $T_N$ of the latter is consistent with its lower $\theta_w$ determined from high

…

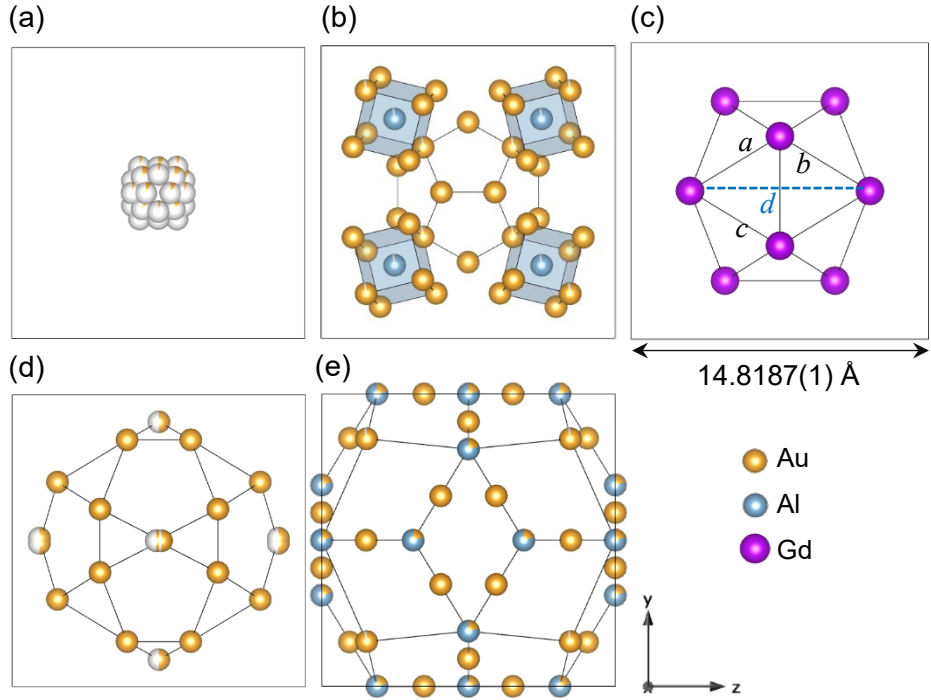

**Fig. 4.** Successive atomic shells of the AAG-II; the innermost unit in (a) is an orientationally disordered tetrahedron. (b) A dodecahedron shell comprised of Au and surrounded by eight cubic units along [111] axes occupied by Al interstices. (c) A Gd icosahedron wherein the *a*, *b*, *c* and *d* distances amount to 5.5337 Å, 5.4785 Å, 5.5337 Å and 8.9629 Å, respectively. (d) Au icosidodecahedron and (e) Au/Al rhombic triacontahedron (RTH).

temperature least-square fitting in Fig. S4 of the Supplemental Material. The insets of Figs. 5(a) and (b) provide temperature-dependence of in-phase component of ac magnetic susceptibility ($\chi'_{ac}$) at frequency ($f$) spanning three orders of magnitude from 0.1 Hz to 100 Hz. Clearly, $\chi'_{ac}$ below $T_N$ in both samples exhibits no discernible frequency dependence, providing strong evidence for the establishment of long-range magnetic order. The absence of frequency dependence in $\chi'_{ac}$ rules out the presence of SG-like dynamics or magnetic frustration, further confirming the stability of the AFM state in both variants.

It is worth noting that positive $\theta_w$ values are commonly reported among AFM QC [6] and ACs such as $Au_{72}Al_{14}Tb_{14}$ 1/1 AC ($\theta_w$ = +4.2 K, $e/a$ = 1.56) [17], $Au_{65}Ga_{21}Tb_{14}$ 1/1 AC ($\theta_w$ = +10.79(2) K, $e/a$ = 1.70) [21,22], $Au_{68}Ga_{18}Dy_{14}$ 1/1 AC ($\theta_w$ = +7.3 K, $e/a$ = 1.64) [26] and $Au_{73}Al_{13}Gd_{14}$ 1/1 AC ($\theta_w$ = +5.9 K, $e/a$ = 1.54) [16], to count a few. The known exceptions exhibiting negative $\theta_w$ include AFM $Cd_6RE$ 1/1 AC series (–32 K < $\theta_w$ < –0.9 K, $e/a$ = 2.14) [14,27] and the AFM $Ga_{50}Pd_{36}Tb_{14}$ 2/1 AC ($\theta_w$ = –10.13 K, $e/a$ = 1.92) [28]. In general, AFM ACs with large $e/a$ ratios close to 2.00 tend to show negative $\theta_w$, whereas those with lower $e/a$ values than 1.70 often exhibit positive $\theta_w$ (see ref. [29] for discussions on this subject). The positive value of $\theta_w$ in these compounds is attributed to their unique magnetic structures. In non-Heisenberg AFM 1/1 ACs (i.e., Tb-, D-, Ho-based ones), the magnetic ordering is of non-coplanar whirling type, stabilized by a dominant FM next-nearest-neighbor interaction ($J_2$) under the condition $J_2/J_1 > 2$ ($J_1$ represents nearest-neighbor interactions). This leads to net FM interaction or positive $\theta_w$. For the Heisenberg-type AFM ACs, there are no experimental reports on their magnetic structures to date.

Figures 5(c) and (d) display the field-dependent magnetization ($M$ versus $H$) at 1.8 K. In both samples, the magnetization saturates at ~ 7 $\mu_B/Gd^{3+}$, i.e., the expected theoretical full-moment value for a free $Gd^{3+}$ ion, above $\mu_0H$ ~ 3 T. The remarkable observation here is the distinct magnetization response of AAG-I and AAG-II to the applied magnetic field; AAG-I exhibits a clear metamagnetic transition at $\mu_0H$ = 0.5 T, while AAG-II does not (it is more visible from the first derivative of $M$ versus $H$). A metamagnetic transition typically corresponds to an abrupt spin reorientation within the AFM phase when a sufficiently strong external magnetic field is applied, leading to a forced FM or other magnetic phase. In contrast, a gradual change in the magnetization of AAG-II suggests a smoother and continuous magnetic field induced spin realignment. This reflects differences in magnetic orders of the two AFM 1/1 AC variants. To rule out the crystallographic orientation of the single crystals as a contributing factor to the presence/absence of the metamagnetic transition, $M$–$H$ curves were collected separately for AAG-I and AAG-II with the magnetic field applied along the [100], [110], and [111] crystallographic axes, as provided in Fig. S5 of the Supplemental Material. The similar magnetization behavior observed along these various orientations confirms that the presence/absence of the metamagnetic transition is an intrinsic property of the samples, rather than an artifact of crystallographic alignment.

…

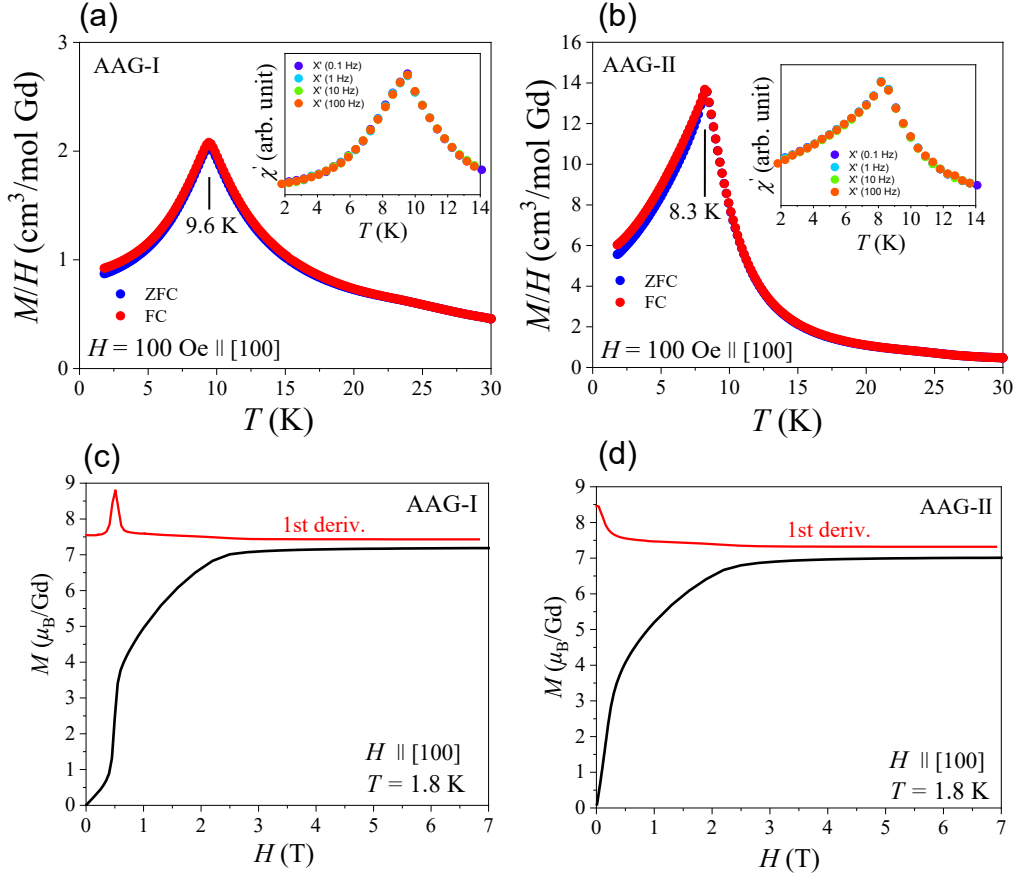

**Fig. 5.** (a) Low temperature magnetic susceptibility $M/H$ of (a) AAG-I, and (b) AAG-II 1/1 AC single crystals. The insets in (a) and (b) show corresponding in-phase component of ac magnetic susceptibility ($\chi'_{ac}$) under $f$ spanning three orders of magnitude from 0.1 to 100 Hz. (c) and (d) provides field-dependent magnetization of AAG-I and AAG-II 1/1 ACs, respectively.

To gain further insight, the specific heat ($C_p$) of the AAG-I and AAG-II was measured under magnetic fields ranging from $\mu_0 H = 0$ to 7 T in 1 T intervals. Figure 6(a) and (b) show the temperature dependence of $C_p/T$ for AAG-I and AAG-II, respectively, under varying magnetic fields. Both samples exhibit a clear sharp jump around their respective $T_N$ confirming AFM order establishment. The anomaly is suppressed above $\mu_0 H \sim 1$ T indicating field-induced perturbation of AFM order. In addition, in both samples, a Schottky-type anomaly appears in $C_p/T$ around $T = 3$ K. This anomaly stems from splitting of energy levels of the $Gd^{3+}$ due to internal magnetic fields, analogous to the Zeeman effect. In a simplified two-level system, Schottky contribution to the specific heat is described by the following equation [30]:

$$C_p(T) = R \left(\frac{\Delta}{kT}\right)^2 \frac{\exp(\Delta/kT)}{(\exp(\Delta/kT)+1)^2} \quad (1)$$

where $\Delta$ is the energy separation between the lowest two states, and $R$ is the gas constant. The low temperature segments of the $C_p/T$ curves in Figs. 6(a) and (b) are best reproduced by assuming energy gaps of $\Delta = 1.60 \times 10^{-22}$ J and $1.35 \times 10^{-22}$ J (corresponding to $\Delta = 8.05$ cm$^{-1}$ and 6.79 cm$^{-1}$, respectively). The larger $\Delta$ in the former suggests that more energy (or a higher temperature) is required for thermal fluctuations to overcome this energy gap and disrupt the AFM order resulting in larger $T_N$, in agreement with experimental results in Fig. 5(a) and (b). It also implies slightly stronger exchange interactions and larger $\theta_w$ for the former, consistent with the experimental data discussed earlier. The Schottky anomaly in both samples gradually shift to higher temperatures with increasing external magnetic field from 0.01 T to 7T, reflecting further field-induced splitting of the energy levels due to Zeeman effect according to $\Delta E = g\mu_B \mu_0 H$, where $g$ denotes $g$-factor. A few examples of compounds showcasing similar Schottky anomaly in their specific heat include spin-glass Zn-Ag-Sc-Tm QC and Zn-Sc-Tm 1/1 AC [31], FM Au-Ge-Gd 1/1 ACs [32], FM GdNi$_5$ [33] and AFM NdFe$_3$(BO$_3$)$_4$ [34].

Figures 6(c) and (d) illustrates temperature dependence of magnetic contribution to the specific heat divided by temperature ($C_M/T$) for the AAG-I and AAG-II, respectively. The $C_M$ is estimated by subtracting nonmagnetic phonon and electronic contributions ($C_{NM}$) from $C_p$ of the compounds ($C_M = C_p - C_{NM}$) using values obtained from an isostructural non-magnetic counterpart with chemical composition Au$_{64}$A$_{22}$Y$_{14}$ 1/1 AC. The insets in Fig. 6(c) and (d) show the variation of magnetic entropy ($S_M$) versus temperature under various magnetic fields within a temperature range of 1.8 K < $T$ < 40 K. Notably, at zero-field, $S_M$ saturates to slightly below $R\ln(2J+1)$ = $R\ln 8$ ($J$ represents the total angular momentum) above $T_N$,

...

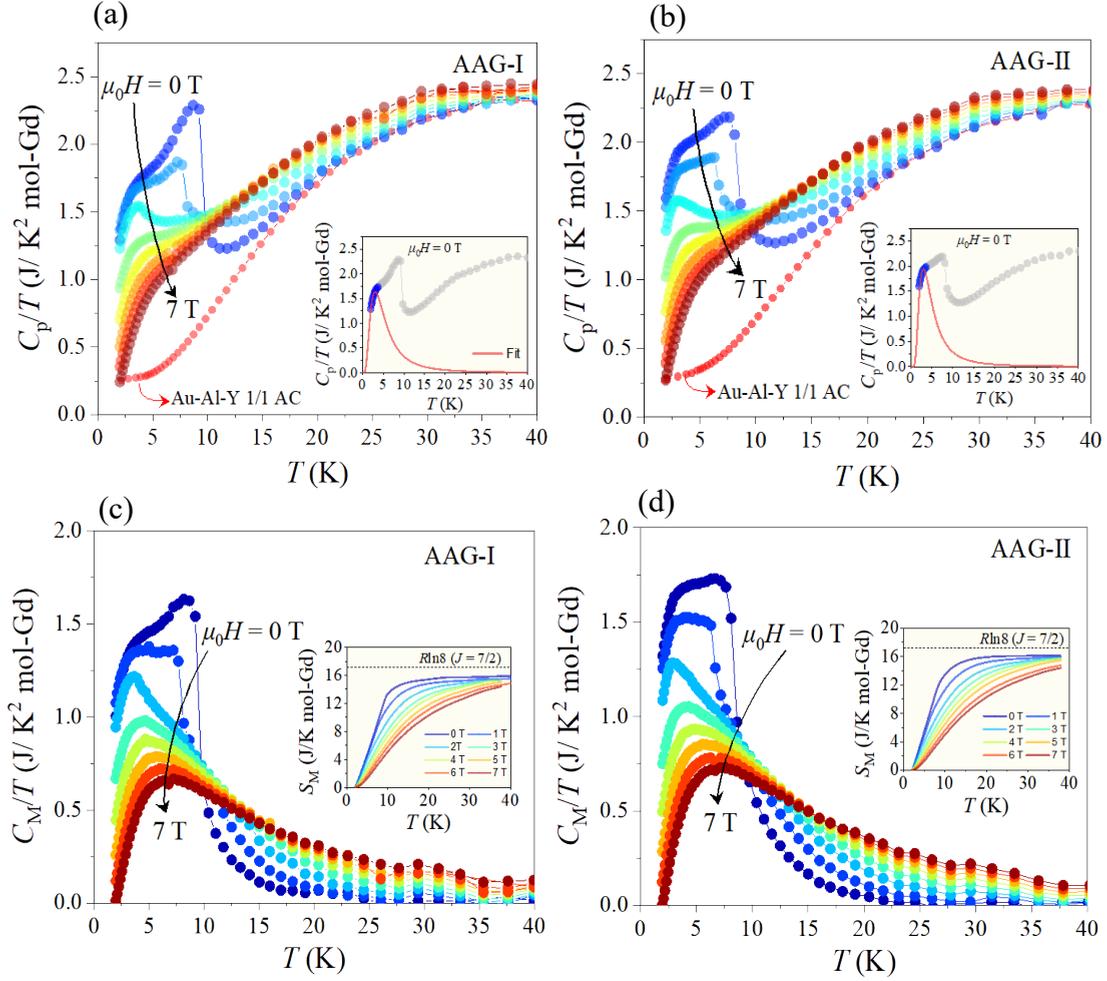

**Fig. 6.** Temperature dependence of $C_p/T$ for (a) AAG-I and (b) AAG-II under different magnetic fields spanning from $\mu_0 H = 0 - 7$ T with 1T intervals. The solid lines in the insets of (a) and (b) represent the Schottky contribution to the specific heat calculated based on equation (1). (c) and (d) show magnetic contribution to the specific heat $C_M/T$ in the corresponding ACs. The insets in (c) and (d) provide $S_M$ versus $T$ estimated by integrating the $C_M/T$ curves under various magnetic fields.

which is close to the expected value for free $Gd^{3+}$ with $J = 7/2$. Note that since our lower temperature limit in specific heat measurement is 1.8 K, the integration does not account for any contributions to the magnetic entropy below this temperature which may be a reason behind slight underestimation of $S_M$ from $R\ln 8$. Below $T_N$, entropy decreases due to spin correlations and AFM order of moments.

As mentioned earlier, the magnetic structures in Tsai-type Heisenberg compounds containing $Gd^{3+}$ ions are not yet well understood, mainly due to the challenges associated with neutron diffraction studies arising from the exceptionally high neutron absorption cross-section of Gd. In this context, theoretical calculations based on realistic atomic structures are essential for narrowing down potential magnetic structures. In one of the latest attempts in this direction, recently [35–38], a comprehensive map of magnetic orders was established for Gd-contained Tsai-type 1/1 ACs through classical Monte Carlo simulations using an effective Heisenberg cluster model with the following Hamiltonian:

$$H = -\sum_{\langle i,j \rangle} J_{\langle i,j \rangle} \boldsymbol{S}_i \cdot \boldsymbol{S}_j - h_z \sum_i S_i^z \qquad (2)$$

where $\boldsymbol{S}_i$ represents the spin vector with unit length at the $i$-th RE site. The term $J_{\langle i,j \rangle}$ represents the RKKY exchange interaction between spins located at the $i$-th and $j$-th RE sites. The magnitude of the applied magnetic field is $h_z$, assuming that the z axis is parallel to the direction of the magnetic field. These simulations with zero magnetic field identified four distinct magnetic phases over a range of Fermi wavenumber ($k_F$) associated with Au concentration (or $e/a$ value): *ferromagnetic*, *incommensurate*, *cluster Néel*, and *cuboc* phases. Among these phases, only the cluster Néel and cuboc are identified as potential commensurate AFM phases.

Given that the primary difference between the two identified AFM phases in the present study lies in the presence or absence of a metamagnetic transition in their magnetization curves, the magnetization curves for both the cluster Néel and cuboc phases are generated numerically by performing variational calculations with adiabatic transformation of the magnetic field

...

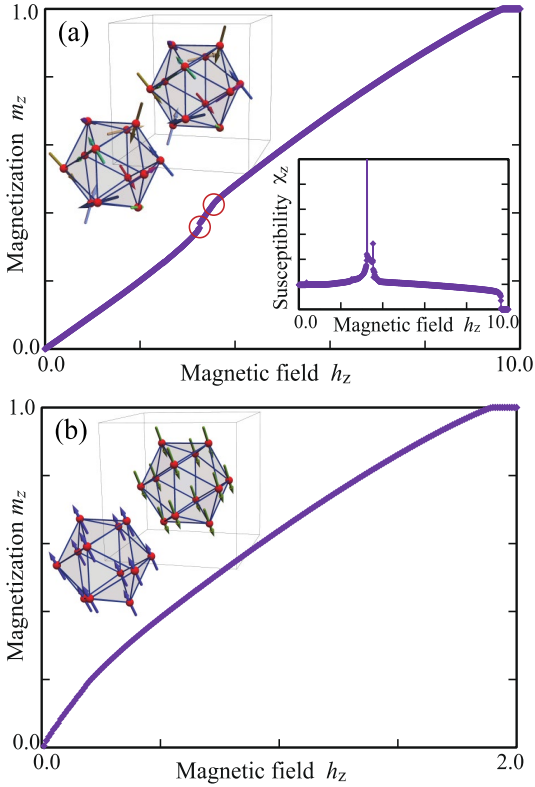

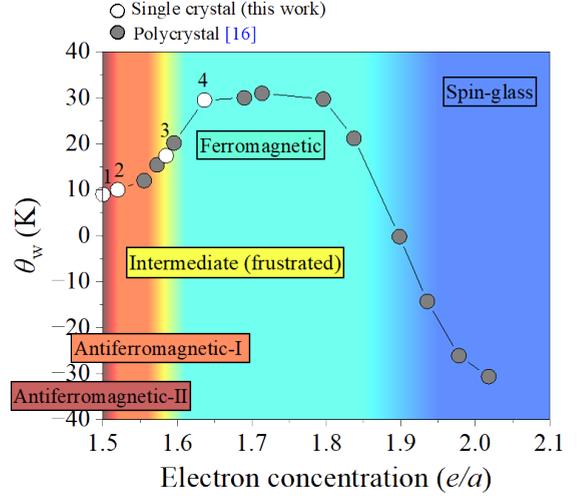

**Fig. 7.** Magnetization curves and zero-field spin configurations for (a) the cuboc and (b) the cluster Néel phases (see [35–38] for the detail) obtained by the variational method, as described in the main text. The normalized magnetization is defined by $m_z = N^{-1}\sum_i S_i^z$, where $N$ is the number of spins in the whole system. In the spin configuration of the cuboc phase, 12 spins in an icosahedron have different directions and the net moment of the icosahedron is zero, while the spin configurations of two icosahedra in a unit cell (cubic corner and body center) are ferroic. On the other hand, in cluster Néel phase, 12 spins in an icosahedron are the same; the net moment is full in an icosahedron. Yet, the net moments of two icosahedra are antiferromagnetic in the cluster Neel phase. The inset of (a) shows susceptibility $\chi_z$, i.e., the field derivative of the magnetization curve $\partial m_z / \partial h_z$.

**Fig. 8.** A magnetic phase diagram of the Au-Al-Gd 1/1 ACs showing $e/a$ dependence of $\theta_w$. The dark blue, light blue, green, yellow and red background colors represent spin-glass, ferromagnetic, intermediate, antiferromagnetic-I and antiferromagnetic-II regimes, respectively. White markers represent the single crystal samples studied in the present work while grey markers correspond to polycrystalline samples reported elsewhere [16].

$h_z$. Starting from the zero-field ($h_z=0$) ground state of an 8×8×8 unit-cell system (12,288 spins), the ground state under applied magnetic fields is determined through a variational optimization using repeated single-update sweeps.

Figure 7 presents the calculated magnetization curves for the (a) cuboc and (b) cluster Néel phases, with insets illustrating schematic representations of the corresponding spin models on icosahedral clusters. As seen in Fig. 7(a), spin-flop and cusp anomalies are observed for the cuboc phase, whereas no such anomalies appear in the cluster Néel phase shown in Fig. 7(b). In fact, the cuboc is the only phase among all theoretically realized ones to generate a metamagnetic anomaly. These behaviors qualitatively agree with the experimental magnetization curves shown in Fig. 5(c) and (d), suggesting that the magnetic structures of the present AAG-I and AAG-II may correspond to the cuboc and cluster Néel phases, respectively. In fact, previous theoretical studies [35–38] support phase transition between these two AFM phases as a result of changes in the $k_F$ (associated with Au concentration) and electron density, which aligns well with the experimental observations in this work. However, the saturation field for the cuboc phase is approximately five times larger than that of the cluster Néel phase. Even at $h_z=4.0$, the cuboc phase does not reach full saturation, while the cluster Néel phase saturates around $h_z=2.0$. Although this difference in saturation fields does not fully match the experimental results in Fig. 5, it could be due to the fact that experimental samples in the present work locate closer to the phase boundary in the magnetic phase diagram, whereas the numerical parameters are set to represent the center of each phase for numerical stability.

Upon identifying and characterizing two distinct AFM orders in the Au-Al-Gd 1/1 ACs, a comprehensive magnetic phase diagram of Tsai-type Heisenberg 1/1 ACs is developed as a function of $e/a$, as illustrated in Fig. 8. Here, the experimental data points from single crystals investigated in the present study (represented by white markers in Fig. 8) are superimposed with data derived from polycrystalline samples reported elsewhere [16] (grey markers). In Fig. 8, AFM phases with and without metamagnetic transitions are labeled as antiferromagnetic-I and antiferromagnetic-II, respectively. Interestingly, upon careful investigation, an intermediate magnetic phase with incommensurate characteristics is identified within a narrow region in between the FM and AFM-I phases at $e/a \approx 1.58$ (highlighted by a yellow region in Fig. 8). In fact, the existence of an incommensurate phase in between FM and AFM phases has been theoretically predicted through Monte Carlo simulations elsewhere [38], though has never been observed experimentally—until now. In the calculations, its appearance is associated with the magnetic frustration between the first neighbor *inter-cluster* interactions and *intra-cluster* interactions [38] (here, intra-cluster interactions refer to magnetic interactions between the moments on the same

…

icosahedron cluster, while inter-cluster interactions involve interactions between adjacent icosahedron clusters).

Figure S6 of the Supplemental Material provides optical microscope images samples number 3 and 4 in a single crystal form. Their corresponding temperature dependence of dc and ac magnetic susceptibilities and inverse magnetic susceptibility are provided in the Supplemental Fig. S7. Sample number 4 clearly exhibits a FM characteristic, evidenced by the sharp rise in $M/H$ and zero frequency dependence of $\chi'_{ac}$ below $T_C = 27$ K. Sharp rise in $M/H$ in FM materials corresponds to the onset of spontaneous magnetization ($M_S$), typically described by $M_S(T) = M_0(-\epsilon)^\beta$, where $M_0$, $\epsilon$ and $\beta$ represents a critical amplitude, a reduced temperature $(T - T_C)/T_C$ and a critical exponent, respectively. Sample number 3, on the other hand, demonstrates a notable frequency-dependent dynamics in $\chi'_{ac}$ below ~ 10 K suggesting metastability in the spin state likely due to enhanced competition between first-neighbor inter-cluster and intra-cluster interactions, as predicted elsewhere [38]. Therefore, as the $e/a$ ratio decreases from values corresponding to the FM region ($1.59 \leq e/a \leq 1.87$), the spin system undergoes a transient meta-stable state at $e/a \approx 1.58$, before reaching AFM-I and subsequently AFM-II states at lower $e/a$ values. Here, for the $e/a$ estimation, valence electron numbers of 1 for Au and 3 for both Al and Gd have been assumed.

The significance of the present results can be viewed in different respects. First, is the experimental discovery of a novel AFM phase (named as AFM-II) in Tsai-type Heisenberg ACs for the first time. Based on the experimental and theoretical investigations, the absence of a metamagnetic transition is a hallmark of this novel AFM variant, which is likely to be a cluster Néel phase, based on the theoretical calculations. This discovery paves the way for exploring even more diverse magnetic states not only in Heisenberg but also in non-Heisenberg members of Tsai-type iQCs and ACs, as predicted theoretically elsewhere [38–41]. Second, the single-crystal synthesis protocol introduced in this study offers a powerful tool for precisely controlling the chemical composition and electron concentration. This level of control enables uncovering magnetic phases that are stable within narrow composition and $e/a$ ranges (such as AFM-II and the intermediate phases in the present work). The advantage of single crystal over polycrystalline sample synthesis in exploring such delicate magnetic phases lies in the fact that the latter typically undergo arc-melting, which can alter the sample weight and lead to slight deviations from the intended chemical composition. Additionally, the single-crystal nature of the samples in this study makes them ideal for detailed investigations of their microscopic nature using techniques such as x-ray resonant magnetic scattering. Finally, a comprehensive magnetic phase diagram developed in this work highlights the rich, tunable magnetic states in Heisenberg Tsai-type ACs. Extending this study to other Tsai-type systems could unveil even more diverse magnetic phases, contributing to the broader field of magnetism in aperiodic structures.

## CONCLUSION

The present study identified two distinct antiferromagnetic (AFM) 1/1 approximant crystals (ACs), designated as AFM-I and AFM-II, in the Au-Al-Gd alloy system, likely corresponding to the cuboc and cluster Néel phases, respectively, based on variational calculations. The AFM-I is distinguished by a field-induced metamagnetic transition in its magnetization curve, while the AFM-II lacks such transition. Moreover, a comprehensive magnetic phase diagram of Heisenberg-type ACs versus electron concentration is established. These findings while advancing our understanding of magnetic phase transitions in these complex systems pave the way for future exploration of more diverse magnetic states in not only Heisenberg but also non-Heisenberg members of the Tsai-type quasicrystal family.

## ACKNOWLEDGMENT

This The authors acknowledge Akiko Takeda for assistance in the synthesis of the materials. The authors also acknowledge Prof. Takenori Fujii in Cryogenic Research Center of the University of Tokyo, Japan, for the heat capacity measurements. This work was supported by Japan Society for the Promotion of Science through Grants-in-Aid for Scientific Research (Grants No. JP19H05817, No. JP19H05818, No. JP21H01044, and No. JP24K17016) and Japan Science and Technology agency, CREST, Japan, through a grant No. JPMJCR22O3.

## RERERENCES

…